\documentclass[prb,twocolumn,showpacs,amsmath,amssymb,floatfix]{revtex4}
\usepackage{graphicx}
\usepackage{dcolumn} 
\usepackage{bm}
\usepackage{amsmath}
\usepackage{latexsym}
\usepackage{amsfonts}
\usepackage{dcolumn}
\usepackage{amssymb}
\usepackage[rightcaption]{sidecap}
\newcommand{\be}{\begin{equation}}
\newcommand{\ee}{\end{equation}}
\newcommand{\bea}{\begin{eqnarray}}
\newcommand{\eea}{\end{eqnarray}}
\newcommand{\nn} {\nonumber}
\renewcommand{\vr} {{\bf r}}

\newcommand{\Tr}{ {\rm Tr} \, }

\def\d{\delta}

\def\l{\lambda}
\def\L{\Lambda}

\def\S{\Sigma}

\def\w{\omega}

\def\xc{{\rm xc}}
\def\x{{\rm x}}

\def\Tr{{\rm Tr}\,}

\begin{document}
%------------------------------------------------------------------------------------------------------------
%--------------------------------------------------Title----------------------------------------------------
%------------------------------------------------------------------------------------------------------------
\title{Correlation energy functional and potential from time-dependent \\
exact-exchange theory}
\date{\today}
\author{Maria Hellgren}
\author{Ulf von Barth}
\affiliation{Department of Mathematical Physics, Institute of Physics, Lund University, 
S\"olvegatan 14A, S-22362 Lund, Sweden}  
\date{\today}
%------------------------------------------------------------------------------------------------------------
%-----------------------------------------------Abstract--------------------------------------------------
%------------------------------------------------------------------------------------------------------------
\begin{abstract}
In this work we have studied a new functional for the correlation energy obtained from 
the exact-exchange (EXX) approximation within time-dependent density functional theory
(TDDFT). Correlation energies have been calculated for a number of different atoms showing
excellent agreement with results from more sophisticated methods. These results loose little 
accuracy by approximating the EXX kernel by its static value, a procedure which enormously simplifies the calculations. The correlation potential, obtained by taking the functional derivative with respect to the density, turns 
out to be remarkably accurate for all atoms studied. This potential has been used to calculate 
ionization potentials, static polarizabilities and van der Waals coefficients with results in 
close agreement with experiment.
\end{abstract}
\maketitle
%------------------------------------------------------------------------------------------------------------
%---------------------------------------------Introduction-----------------------------------------------
%------------------------------------------------------------------------------------------------------------
\section{Introduction}
Time-dependent density functional theory (TDDFT) provides a promising
and rigorous framework for treating interacting many-electron systems
within a reasonable computational cost. 
Although the overall aim of TDDFT is to describe physical phenomena
associated with excited states, strong connections between the latter
and ground-state properties make it possible to obtain an improved 
description of the ground state through the use of TDDFT.

In many interesting cases, it is sufficient to treat systems under the
influence of external perturbations weak enough to allow for a 
description in terms of linear response. Within this realm, the basic
quantities of TDDFT are the ground-state exchange-correlation (XC)
potential and the corresponding XC kernel ($f_\xc$).\cite{gk85} In previous 
work we have studied these quantities at different levels
of approximation. The XC potential has been calculated for atoms at
the level of the $GW$ approximation\cite{hvb07} and the XC kernel 
has been presented within the exact-exchange (EXX) approximation.\cite{hvb08,hvb09,kvb}
In all these previous publications the necessary formulas have been
derived from the variational formulation of many-body perturbation
theory.\cite{vbdvls05,abl99} The virtue of this approach is that the obtained results are
guaranteed to obey many conservation laws and sum rules like, e.g.,
the $f$-sum rule, of importance to the calculated optical spectra.
In the present work we have deviated from this path and instead
followed an approach originally suggested by Peuckert.\cite{p78} 

Having some approximation for the density-density response function 
of a many-electron system one can obtain an expression for the total
ground-state energy by using the Hellmann-Feyman theorem applied to
the strength of the inter-particle Coulomb interaction.\cite{coaun,lp75} 
And, by means of TDDFT, from any approximation for the XC potential and the
corresponding XC kernel one can obtain an approximation to the density
response function of the system. The XC part of the resulting total
energy can then be differentiated once with respect to the density to
yield a new approximation for the XC potential and then twice to yield a
new approximation for the XC kernel. From these results we can obtain
a new total energy which again can be differentiated to obtain a new
potential and a new kernel, and so on. Of course, in practice the
resulting expressions quickly become unmanageable and the proposed
iterations have to be limited to one or two steps.

In previous work we have seen that the total energy obtained via the
Hellman-Feynman theorem applied to the EXX approximation is very 
accurate giving errors of the order of 5\% in the resulting correlation
energies.\cite{hvb08} These results inspired us to believe that a differentiation
of the corresponding expression for the XC energy with respect to the
density might give rise to a very accurate XC potential and XC kernel. 
And this is, indeed, what we have found in the
present work, at least as far as the potential is concerned and, then, 
within the approximations we have been forced to do in order to
obtain tractable expressions.

In the present context, the XC kernel of the EXX approximation (EXXA)
has the convenient property of being linear in the Coulomb interaction
allowing us to carry out the integration over the strength of the
Coulomb interaction analytically. The result is a closed expression
for the XC energy. This expression explicitly contains the EXX kernel
$f_\x$, which give rise to numerical difficulties in the later process
of differentiating the XC energy with respect to the density. Motivated
by the fact that the EXX kernel for He is independent of the density
we have neglected this density dependence for
all atoms. The validity of this approximation must, of course, be
verified independently but, encouraged by our excellent results, we
have decided to leave this test to a future publication.

We have also employed an additional approximation which we believe to be
of even less consequence. The fact that the EXX kernel is frequency 
dependent for all systems but for a two-electron one leads to much
longer computational times. We have, however, seen that total energies
are very insensitive to this frequency dependence and we thus recommend to
neglect it. The frequency dependence will, of course, also affect the
calculation of the XC potential except in the case of He. We have here
assumed that also the potentials of the heavier atoms are relatively
insensitive to this frequency dependence but we leave also the
verification of this assumption to a future publication. 

The XC potentials that we obtain from the first iteration of the
Peuckert procedure starting from the EXXA turns out to be much better
than any approximate XC potentials perviously obtained and they are
actually very close to the exact ones\cite{ug94} where these are known. For a
number of spherical atoms, we have used these new potentials to
calculate total energies, static polarizabilities, and van der Waals
coefficients using the expression for the XC kernel obtained within the
EXXA. In all cases we find excellent agreement with experiment. We thus
conclude that we now have an affordable and well defined way of
obtaining accurate results for ground-state properties and low lying
excitations of many-electron systems.    
%------------------------------------------------------------------------------------------------------------
%---------------------------------------------Introduction-----------------------------------------------
%------------------------------------------------------------------------------------------------------------
\section{Correlation energy functional}
A standard expression for the correlation energy can be obtained by introducing 
a fictitious Hamiltonian $H^\l$ with a scaled Coulomb interaction $\l v$ and a local 
multiplicative potential which guarantees that the density is constant for every value 
of the scaling parameter $\l$. At $\l=1$, $H^\l$ coincides with the fully interacting 
Hamiltonian and at $\l=0$ with the one of the non-interacting Kohn-Sham (KS) system. 
Using the Hellman-Feynman theorem one can show\cite{coaun,lp75,tddftbook} 
\be
E_{\rm c}=\frac{i}{2}\int_0^1 d\lambda \int\frac{d\omega}{2\pi}\,\,{\rm Tr}\{v[\chi^{\lambda}(\omega)-\chi_s(\omega)]\},
\label{corr}
\ee
where $\chi_s$ is the non-interacting KS response function and $\chi^{\l}$ is the scaled density response function. 
We have also used the short hand notation $\Tr fg=\int d{\vr}d{\vr'} f({\vr},{\vr'})g({\vr'},{\vr})$ for 
any two-point functions $f$ and $g$. 
Within TDDFT the function $\chi^{\l}$ reads 
\be
\chi^{\l}=\chi_s+\chi_s\left[\l v+f^{\l}_\xc\right]\chi^{\l}.
\label{rpafx}
\ee
The scaled XC kernel $f^\l_{\xc}$ is a functional of the ground-state density and is defined 
as the functional derivative of the scaled XC potential $v^\l_{\xc}$ with respect to the 
density $n$. 

The simplest approximation to Eq. (\ref{corr}) is the random phase approximation (RPA), obtained with $f^\l_{\xc}=0$. The RPA has the advantage of allowing for an analytical evaluation of the $\l$-integral in Eq. (\ref{corr}).
The result is
\be
E_{\rm c}=-\frac{i}{2} \int \frac{d\omega}{2\pi}\,\,\Tr\{\ln[1-v\chi_s]+v\chi_s\}.
\label{ec}
\ee
In the language of Feynman diagrams Eq. (\ref{ec}) is equal to an infinite summation 
of ring-diagrams. 

The RPA correlation potential $v_{\rm c}$ is obtained as the functional derivative of Eq. (\ref{ec}) 
with respect to the density. If we let $V$ signify the total KS potential,  $G_s$ the non-interacting 
KS Green function, $\chi_s=-iG_sG_s$ and the functional derivative is 
conveniently obtained via the chain rule 
$$
\frac{\d E_{\rm c}}{\d n}\frac{\d n}{\d V}=\frac{\d E_{\rm c}}{\d G_s}{\frac{\d G_s}{\d V}}.
$$ 
The result is the well-known linearized Sham-Schl\"uter (LSS) equation\cite{ss83,vbdvls05}
\be
\int\chi_s(1,2)v_{\rm c}(1)d2=\int \L(3,2;1)\S_{\rm c}(2,3)d2d3.
\label{lss}
\ee
Here, we have used the notation $(\vr_1,t_1)= 1$ etc. and introduced
$\L(3,2;1)=-iG_s(3,1)G_s(1,2)$. The correlation part of the self-energy $\S_{\rm c}$ in the RPA 
is given by
\be
\S_{\rm c}=i\frac{\delta E_{\rm c}}{\delta G_s}=iv\chi^{\rm RPA} vG_s
\ee
where 
\be
\chi^{\rm RPA}=\chi_s+\chi_sv\chi^{\rm RPA}.
\label{rpa}
\ee
Thus, the $\S_{\rm c}$ in the RPA coincides with the $GW$ self-energy but evaluated at KS Green functions.
For the purpose of obtaining the RPA potential Eq. (\ref{lss}) and Eq. (\ref{rpa}) have to be solved self-consistently and has only been done so far for atoms\cite{hvb07} and in bulk Si, LiF and solid Ar.\cite{gmr06}
The RPA is also sometimes called the linearized time-dependent (TD) Hartree approximation 
since $\chi^{\rm RPA}$ 
is obtained by allowing the electrons to respond only to the perturbing potential plus the 
induced Hartree potential. The next level of approximation is obtained by also including exchange effects, which leads to the RPAE approximation or the linearized TD Hartree Fock (TDHF) approximation. Within TDDFT the same level of approximation corresponds to the TDEXX approximation, in which the HF potential is replaced by the local EXX potential $v_\x$. The latter potential is obtained from the TD exchange version of the 
LSS equation which amounts to replacing, in Eq. (\ref{lss}), $v_{\rm c}$ by $v_\x$ and $\S_{\rm c}$ by $\S_\x=ivG_s$, i.e., the HF self-energy. A variation of that equation yields an equation for the EXX response kernel $f_\x$:
\begin{eqnarray}
&&\int \chi_s(1,2)f_{\x}(2,3)\chi_s(3,4)d2d3\nn\\
&&\,\,\,\,\,\,\,\,\,\,=\int \frac{\delta\S_\x(2,3)}{\delta V(4)}\Lambda(3,2;1)d2d3\nn\\
&&\,\,\,\,\,\,\,\,\,\,\,\,\,\,+\int \Lambda(1,2;4)\Delta(2,3)G_s(3,1)d2d3\nn\\
&&\,\,\,\,\,\,\,\,\,\,\,\,\,\,+\int G_s(1,2)\Delta(2,3)\Lambda(3,1;4)d2d3,
\label{fxceq}
\end{eqnarray}
where $\Delta(2,3)=\S_\x (2,3)-v_{{\x}}(2)\delta(2,3)$. A full analysis of this kernel has been 
performed recently.\cite{hvb09}
It is important to observe that the kernel $f_\x$ is an implicit functional of the density through the 
KS orbitals and their eigenvalues. From Eq. (\ref{fxceq}) we see that the potential $v_\x$ is an 
ingredient in the kernel $f_\x$ but, as discussed above, $v_\x$ is also 
an implicit functional of the density. For future reference we here note that an EXX kernel obtained for any density and used in Eq. (\ref{rpafx}) will generate an interacting response function which obeys the important $f$-sum rule. This is due to the fact that the procedure above guarantees that the kernel
does not blow up at large frequencies, something that we proved in a previous publication.\cite{hvb08} 
The fulfillment of the $f$-sum rule is crucial for the construction of new functionals for the correlation energy, one of which we will now derive.

We start by making the observation that the EXX kernel is linear in the explicit dependence on the Coulomb potential. Therefore, the $\l$-integration in Eq. (\ref{corr}) can, just as in the case of RPA, be carried out analytically yielding the result 
\be
E_c=-\frac{i}{2} \int \frac{d\omega}{2\pi}\,\,\Tr\left\{\frac{v}{v+f_\x}\ln[1-\left[v+f_\x\right]\chi_s]+v\chi_s\right\}.
\label{ecfx}
\ee
We note that $\Tr v\chi_s$ contains a singularity proportional to the Coulomb potential at the origin times the number of particles. This singularity must be cancelled by the first term in Eq. (\ref{ecfx}) which will  only occur if the kernel remains finite at large frequencies, i.e., obey the sum rule. 

\begin{figure}[t]
\includegraphics[width=8.5cm, clip=true]{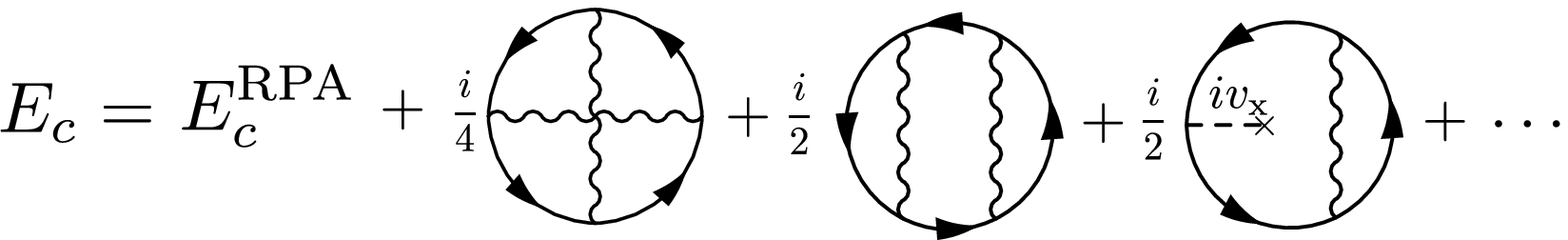}\\
\caption{With $f_\xc=f_{\rm x}$, the correlation energy functional has a diagrammatic expansion up to second order. Higher order terms are simulating higher order exchange diagrams.}
\label{ediag}
\end{figure}

Correlation energies obtained from this functional was recently presented\cite{hvb08} but in a non self-consistent fashion using the EXX density for the evaluation. The results were excellent for many closed-shell atoms. A diagrammatic description of Eq. (\ref{ecfx}), up to second order, is given in Fig.~\ref{ediag}. Many higher order terms do not have a strict diagrammatic representation but can be considered to simulate the higher order exchange diagrams. 

In order to obtain the correlation potential from Eq. (\ref{ecfx}) we need to differentiate the correlation energy with respect to the density. We then encounter the problem of differentiating the EXX kernel 
with respect to the density, which is a rather cumbersome task. Such derivatives formally amounts to 
three-point correlation functions $\d f_\x(1,2)/\d n(3)$. For He, however, this derivative is zero since $f_{\rm x}=-\frac{1}{2}v$. Therefore, we have here chosen to see how well we can do by neglecting the density variation of the EXX kernel also for larger atoms. 
With this assumption the same procedure used for going from Eq. (\ref{ec}) to Eq. (\ref{lss}) now leads to an equation similar to Eq. (\ref{lss}) with $\S_{\rm c}$ replaced by 
\be
\S_{\rm c}=iv\chi\left[v+f_\x\right]G_s,
\ee
where $\chi$ is the interacting response function in the TDEXX approximation (Eq. \ref{rpafx} at $\l =1$ and $f_\xc=f_\x$).  
The evaluation of the correlation energy form Eq. (\ref{ecfx}) is relatively time consuming due to the frequency dependence of the kernel $f_\x$.\cite{hvb08} The computational cost would be substantially reduced it the kernel could be kept at its static value ($f_\x(\vr,\vr',\w)\approx f_\x(\vr,\vr',0)$) without seriously affecting the correlation energies. In Table \ref{cornrgy} the correlation energies evaluated using the fully frequency-dependent EXX kernel (TDEXX) are compared to those obtained by using the static (=adiabatic) kernel (AEXX) and the difference is seen to be relatively small. Notice that in the case of He the static kernel is the full kernel. 
These results have encouraged us to evaluate the correlation potentials also using the adiabatic approximation.
%------------------------------------------------------------------------------------------------------------
%------------------------------------------------Results--------------------------------------------------
%------------------------------------------------------------------------------------------------------------
\section{Correlation energies and potentials}
When evaluating the correlation energy for a system it is natural to use the self-consistent 
density for that particular system. From now on, the self-consistent density corresponding to the new functional in Eq. (\ref{ecfx}) is referred to as the RPAX density. 
Due to the stationary property of the total energy we expect, however, that evaluating correlation 
energies at a slightly different density will give almost the same result.\cite{foot1} This is, indeed, 
the case as can be seen in Table~\ref{cornrgy}, which also seems to demonstrate that the total 
energy is a minimum at the RPAX density. 
Clearly, the EXX kernel gives a large improvement over the too large RPA values. It also improves the MP2 results, which are here the 
self-consistent results given in Ref. \onlinecite{je05}. The latter approximation also follows 
from Eq. (\ref{corr}) if $\chi^\l$ is replaced by $\chi^\l\approx \chi_s+\l\chi_s\left[v+f_\x\right]\chi_s$, or if the logaritm in Eq. (\ref{ecfx}) is expanded to second order. 
The use of the adiabatic approximation is seen to be less severe yielding energies of the same 
quality as those obtained with the frequency-dependent kernel. As noticed before, the values in TDEXX, or in AEXX for that matter, are very accurate for these systems. 
\begin{table}[t]
\caption{Correlation energies from a few different approximations. For a consistent comparison with CI results the correlation energy is here defined as the difference between the total energy and the Hartree-Fock energy. The fourth decimal is shown in parentheses in order to compare different approximations. (a.u.)}
\begin{ruledtabular}
\begin{tabular}{c|llll|l|l|l}
\multicolumn{1}{l|}{$f_\xc$:}&TDEXX&AEXX&TDEXX&AEXX&RPA&MP2\footnotemark[2]&CI\footnotemark[3]\\
\multicolumn{1}{l|}{$v_\xc$:}&EXX&EXX&RPAX&RPAX&RPA&MP2&\\ [0.5ex]\hline
He&0.044(5)&0.044(5)&0.044(6)&0.044(6)&0.083&0.047&0.0420\\[0.5ex]
Be&0.102(0)&0.101(7)&0.103(3)&0.102(8)&0.181&0.124&0.0943\\[0.5ex]
Ne&0.388(9)&0.377(1)&0.390(3)&0.377(8)&0.596&0.480&0.3905\\[0.5ex]
Mg&0.445(2)&0.437(2)&0.446(6)&0.438(4)&0.681&0.514&0.4383\\[0.5ex]
Ar&0.727(8)&0.710(6)&0.728(7)&0.711(2)&1.091&0.844&0.7225\\[0.5ex]
\end{tabular}
\end{ruledtabular}
\label{cornrgy}
\footnotetext[2]{From Ref. \onlinecite{je05}.}
\footnotetext[3]{From Ref. \onlinecite{cgdpf93}.}
\end{table} 

\begin{figure}[b]
\includegraphics[width=8.5cm, clip=true]{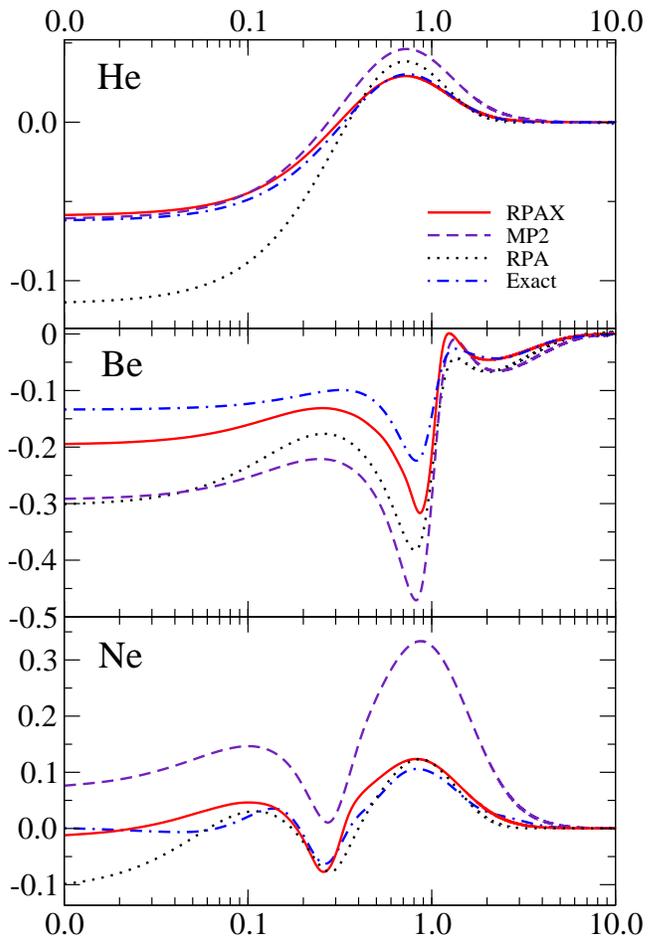}\\
\caption{Self-consistent correlation potentials for He, Be and Ne in different approximations. The MP2
potential for Be is evaluated at the EXX density due to a variational instability.\cite{je05} The exact potentials are those of Umrigar et al.\cite{ug94}}
\label{potentialer}
\end{figure}

The XC potential is interesting in its own right. The highest occupied eigenvalue of the KS system 
exactly corresponds to the ionization potential\cite{avb85} and the larger part of the particle-conserving excitation energies consists of KS eigenvalue differences.\cite{pgg96} 
In Tab.~\ref{ionnrgy} we present ionization potentials produced by different KS potentials 
(RPAX, RPA, MP2 and EXX). As noticed previously, the RPA values improve over EXX 
and are also better than the MP2 values. Here, we see that the RPAX 
potential yields an even further improvement giving excellent ionization potentials for all atoms. 
In Fig.~\ref{potentialer} we plot the same correlation potentials for He, Be, and Ne. 
The He RPAX potential almost coincide with the exact potential\cite{ug94} and 
the RPAX potential for Be is much closer to the exact than any other approximation we 
have tried - especially in the outer region. The Ne potential is also very accurate, yielding even a qualitative improvement by better describing the 2s-shell giving rise to an extra structure in the RPAX, and the exact, correlation potential. 
\begin{table}[b]
\caption{Ionization potentials obtained from the highest occupied KS eigenvalue of different KS potentials. (a.u.)}
\begin{ruledtabular}
\begin{tabular}{llllll}
Atom&EXX&MP2&RPA&RPAX&Exp.\\[0.5ex]\hline
He&0.918&0.893&0.902&0.904&0.904\\[0.5ex]
Be&0.309&0.357&0.354&0.340&0.343\\[0.5ex]
Ne&0.851&0.657&0.796&0.787&0.792\\[0.5ex]
Mg&0.253&0.302&0.297&0.282&0.281\\[0.5ex]
Ar&0.591&0.558&0.590&0.577&0.579\\[0.5ex]
\end{tabular}
\end{ruledtabular}
\label{ionnrgy}
\end{table} 
\section{Static polarizabilities and van der Waals coefficients}
The static polarizability is defined according to the formula
\be
\alpha(0)=-\int z\chi(\vr,\vr',\w=0)z'd\vr d\vr',
\label{pol}
\ee
and the van der Waals coefficient, or $C_6$-coefficient, between ion $A$ and $B$ is  given 
by 
\be
C_6=\frac{3}{\pi}\int_0^\infty \alpha_A(i\w)\alpha_B(i\w)d\w,
\label{c6}
\ee
\begin{table}
\caption{Static polarizabilities calculated from $\chi$ in the RPA and the TDEXX approximation. The latter has been evaluated using different potentials (EXX, RPA, RPAX, and Exact). (a.u.)}
\label{statpolvan}
\begin{ruledtabular}
\begin{tabular}{c|cccc|c|c}
\multicolumn{1}{l|}{$f_\xc$:}&\multicolumn{4}{l|}{TDEXX}&RPA&Litt.\\
\multicolumn{1}{l|}{$v_\xc$:}&\text{EXX}&RPA&RPAX&Exact&RPA&\\[0.5ex] \hline
He&1.322&1.351&1.348&1.349&1.225&1.38\footnotemark[1]\\[0.5ex]
Ne&2.372&2.577&2.613&2.555&2.424&2.67\footnotemark[1]\\[0.5ex]
Ar&10.74&10.69&10.94&-&9.839&11.08\footnotemark[1]\\[0.5ex]
Be&45.64&40.09&41.04&40.49&28.99&37.7\footnotemark[2]\\[0.5ex]
Mg&81.66&70.37&71.67&-&51.56&71.35\footnotemark[2]
\end{tabular}
\label{statpol}
\footnotetext[1]{From Ref. \cite{hibg05}}
\footnotetext[2]{From Ref. \cite{mb03}}
\end{ruledtabular}
\end{table}
\begin{table}
\caption{van der Waals or $C_6$ coefficients calculated from $\chi$ in the RPA, the AEXX and the TDEXX approximation. The latter has been evaluated using different potentials (EXX, RPA, RPAX, and Exact). (a.u.)}
\begin{ruledtabular}
\begin{tabular}{c|cccc|c|c|c}
\multicolumn{1}{l|}{$f_\xc$:}&\multicolumn{4}{l|}{TDEXX}&\multicolumn{1}{l|}{AEXX}&RPA&Litt.\\
\multicolumn{1}{l|}{$v_\xc$:}&\text{EXX}&RPA&RPAX&Exact&RPAX&RPA&\\ [0.5ex]\hline 
He&1.375&1.414&1.411&1.411&1.411&1.206&1.458\footnotemark[1]\\[0.5ex]
Ne&5.506&6.091&6.191&6.021&6.161&5.523&6.383\footnotemark[1]\\[0.5ex]
Ar&61.88&61.27&63.19&-&63.11&53.69&64.3\footnotemark[1]\\[0.5ex]
Be&282.8&226.7&235.3&231.5&236.5&142.0&214\footnotemark[2]\\[0.5ex]
Mg&767.5&617.8&634.3&-&632.3&385.6&627\footnotemark[2]
\end{tabular}
\label{statpolvan}
\footnotetext[1]{From Ref. \onlinecite{km85}.}
\footnotetext[2]{From Ref. \onlinecite{pd02}.}
\end{ruledtabular}
\end{table}
where $\alpha_A(i\w)$ is the dynamic polarizability of ion $A$ calculated at imaginary frequencies. 
In previous work these quantities were calculated using the response function of the TDEXX approximation. It was then a natural choice to use the corresponding self-consistent EXX density in the evaluation. We then found, not so surprisingly, that our results closely resembled those of the TDHF approximation and were, therefore, not overly impressive. We interpreted this partial failure 
to be a consequence of a rather poor description of the ground state. We found, however, also that the results are rather sensitive to the density used in the evaluation. In the present work we have instead evaluated the same EXX formula for the polarizability using the correlated density produced by our new correlation functional. The van der Waals energy is a pure correlation effect and one can argue that it should be evaluated at a correlated density and not just the exchange density like, e.g., the RPA or the RPAX density. This turns out to have rather drastic effect on the actual values moving them much closer to the more accurate values found in the literature as seen in Tab. \ref{statpol} and \ref{statpolvan}. 
A clear improvement is found when correlated densities are used in the evaluation with a slight 
edge for our new RPAX density. It is also noticed that when calculating $C_6$-coefficients the adiabatic approximation is sufficient as also observed previously.\cite{shh06}
%------------------------------------------------------------------------------------------------------------
%---------------------------------------------------C&D--------------------------------------------------
%------------------------------------------------------------------------------------------------------------
\section{Conclusions and Outlook}
In the present work we have decided to temporarily part from our 
familiar, systematic and conserving way of constructing improved
approximations to XC potentials and kernels within TDDFT in the
linear regime. Instead, we have tested the first step in an 
iterative scheme originally proposed by Peuckert. The starting
point has been the previously studied EXX approximation for the
XC potential and corresponding kernel. The method is described in
detail in the sections above. The rational for this deviation has
been the extraordinary accurate potentials obtained from this 
approach. The potentials are indeed very close to the exact XC
potentials of DFT where these are known. Calculated ionization
potentials for all atoms are close to experimental results in
accordance with the well known fact that the highest occupied DF
eigenvalue should equal minus the ionization potential.

We have in previous work calculated static polarizabilities, low
lying particle-hole excitation energies, and van der Waals 
coefficients from the EXX approximation. These turned out to be very
similar to those of TDHF theory and not very
accurate. We argued then that these less satisfactory results were
a consequence of a rather poor description of the ground state within
the EXX approximation. Using instead, our new potentials to recalculate
the mentioned properties results are in excellent agreement with experiment
for all atoms studied.

We can thus state with confidence that we now have at our disposal an
affordable and not too complicated way of obtaining accurate results
for a large number of static and low-frequency dynamic properties of
physical systems. And all this within a relative minor modification
of the EXX approximation within TDDFT.

Of course, we have, at this stage, no idea about the possible 
conserving properties of the proposed scheme. This is an interesting
project for future research. But, on the other hand, conservation
laws and sum rules might not be of such great importance to the
properties discussed here. 

We should also mention that the results obtained here rely on the
assumption that one can neglect the density and the frequency
dependence of the XC kernel of the EXX. Based on the fact that this is
no assumption in the case of a two-electron system we believe in
the validity of this assumption but this must, of course, be further
investigated. In the meanwhile we propose to use the new method as
an effective tool for calculating physical properties within the realm
of TDDFT.
 \begin{acknowledgments}
This work was supported by the European Theoretical Spectroscopy 
Facility (INFRA-2007-211956).
\end{acknowledgments}


\begin{thebibliography}{24}
\bibitem[{foo()}]{foot1}
\bibinfo{note}{We notice that the presented correlation energies are not
  obtained by just evaluating Eq. (8) at different densities but are calculated
  as the total density functional energy minus the exact HF energy.}
\end{thebibliography}
\end{document}